\documentclass[11pt]{article}

\usepackage{epsfig,latexsym}
\usepackage{amsmath}
\usepackage[english]{babel}
\usepackage{graphicx}
\usepackage{lineno}

\newcommand{\qed}{\hfill $\Box$}

\newtheorem{prop}{Proposition}

\title{A note on hospital financing: \\
local financing vs. central financing}

\author{Raffaele Mosca}

\begin{document}

\maketitle


\begin{center}
{\footnotesize
Dipartimento di Economia, Universit\'a degli Studi "G. d'Annunzio", Pescara 65127, Italy.\\
\texttt{raffaele.mosca@unich.it} }
\end{center}




\begin{abstract}

This note tries to study how hospital behaviors, with reference to interhospital collaboration or competition, could be affected by hospital financing systems.

For that this note simulates two scenarios which start with the following baseline scenario: a State, with a set of hospitals, each with all types of wards at a basic level. The evolution of this baseline scenario consists in the evolution of hospitals, that is, in the possibility of hospitals to make some of their wards excel. The State has a budget, for the evolution of this baseline scenario, which can be used by two financing systems: either by a "local financing", i.e., by splitting the budget among the hospitals so that each hospital is managing its own portion of the budget by pursuing the individual benefit, or by a "central financing", i.e., by not splitting the budget among the hospitals so that the State is the sole manager of the budget, by pursuing the benefit of the whole community.

The conclusions seem to be that: in the local financing system hospitals tend to diversify their excellences, while in the central financing system the State tends to create poles of excellence.

\end{abstract}


\section{Introduction}

This note is inspired to reference \cite{[a7]} where the authors study interhospital collaboration and competition. In particular, interhospital collaboration seems to rely on patient transfers, while interhospital competition seems to rely on ability to attract patients. Such aspects in turn seem to rely on the "level difference", i.e. on the suitability difference,  between pairs of hospitals.

Then this note tries to study how hospital behaviors, with reference to such aspects, could be affected by hospital financing system: in fact, the assumption is that the "level difference" between pairs of hospitals in turn relies on the health policy of making hospital wards excel, which finally relies on hospital financing systems.

For that this note simulates two scenarios which start with the following baseline scenario: a State, with a set of hospitals, each with all types of wards at a basic level. The evolution of this baseline scenario consists in the evolution of hospitals, that is, in the possibility of hospitals to make some of their wards excel. The State has a budget, for the evolution of this baseline scenario, which can be used by two financing systems: either by a "local financing", i.e., by splitting the budget among the hospitals so that each hospital is managing its own portion of the budget by pursuing the individual benefit, or by a "central financing", i.e., by not splitting the budget among the hospitals so that the State is the sole manager of the budget, by pursuing the benefit of the whole community.


In details:

- Section 2 introduces the above baseline scenario and the above financing systems (local financing, central financing) according to which this baseline scenario can evolve;

- Section 3 studies the local financing system as a Game Theory problem;

- Section 4 studies the central financing system as a Mathematical Programming problem.

The conclusions seem to be that: in the local financing system hospitals tend to diversify their excellences, while in the central financing system the State tends to create poles of excellence.

Let us mention that the question "centralization or decentralization" in health policy is treated in \cite{[Book]} (cf. Chapter 3) in terms of decision-making and in more general terms, as well as in different papers \cite{[Ref.1],[Ref.2],[Ref.3],[Ref.4],[Ref.5],[Ref.6]} in different terms.

Let us observe finally that similar conclusions could be reached for similar contexts, e.g., for university financing.

\section{Baseline scenario}

Assume that the baseline scenario is defined as follows:

\begin{itemize}
\item[(1.1)] the context is a $State$;

\item[(1.2)] there is a set $Q = \{q_1, q_2,\ldots,q_{|Q|}\}$ of $hospitals$;

\item[(1.3)] set $Q$ denotes, at the same time, the set of $districts$ in which the State is consequently partitioned [one district for one hospital]; in particular set $Q$ is associated with a population distribution $(a_1, a_2, \ldots,a_{|Q|})$ where component $a_i$ denotes the population fraction of district $q_i$ (for $i = 1,\ldots, |Q|$);

\item[(1.4)] there is a set $R = \{r_1, r_2,\ldots,r_{|R|}\}$ of (types of) {\em wards};

\item[(1.5)] each hospital has got all (types of) wards; in particular, all wards of each hospital are of the same level, we say that they are {\bf basic} wards.
\end{itemize}

Let us assume that the evolution of this baseline scenario consists in the evolution of hospitals which is defined as follows: the evolution of a hospital consists in making wards excel, i.e., in transforming some of its basic wards into {\bf excellent} wards.

Such potential excellent wards are aimed at a "new market" of patients [which live in the State], formed by those patients who would be either transferred outside the State or directly hospitalized outside the State, since doctors or patients themselves would assume that State hospitals' basic wards are not suitable.    




Then let us assume that the "new market" is defined as follows:

\begin{itemize}
\item[(1.6)] the "new market" is a set $P$ of $patients$ that, according to an estimation, in a certain time frame should be hospitalized in a potential excellent ward of type $r$ (for $r = 1,\ldots,|R|$);

\item[(1.7)] set $P$ is partitioned into groups $P_1,P_2,\ldots, P_{|R|}$, where $P_r$ is the group of patients that, according to an estimation, in a certain time frame should be hospitalized in a potential excellent ward of type $r$ (for $r = 1,\ldots,|R|$); patients of group $P_r$ (for $r = 1,\ldots,|R|$) are distributed over districts according to population distribution $(a_1, a_2,\ldots, a_{|Q|})$.
\end{itemize}

Then let us assume that the State has got a "budget" to finance the evolution of hospitals and that such a budget can take place by one of the following financing systems:

- {\em local financing} : the State splits the budget among the hospitals, each of which is managing its own portion of the budget. Each hospital will pursue the individual benefit: this benefit is defined as maximizing the number of patients of the "new market" which will be hospitalized in that hospital.

- {\em central financing} : the State is the sole manager of the budget. The State will pursue the whole community benefit: this benefit is defined as minimizing the sum of costs to make the chosen wards (of the chosen hospitals) excel, plus the sum of costs/discomforts that patients of the "new market" should incur to be hospitalized (either inside or outside the State).

Then let us try to study the evolution of this baseline scenario for such financing systems.

\section{Local financing}

The study of local financing is inspired to reference \cite{[a12]}. For that, first we report a part of reference \cite{[a12]} in which the authors study a scenario as a Game Theory problem, then we study local financing as a similar Game Theory problem.

For any reference about Game Theory let us refer to \cite{[Osborne]}. Let us recall that a {\em game} is a scenario in which individuals (players) can choose respectively certain behaviors (strategies), with the aim to maximize respectively their benefits (payoffs), taking into account that their respective payoffs depend on the strategies chosen by all players. The goal of Game Theory is to predict, given a game and given certain assumptions on players' rationality, players' behavior and therefore a game outcome. In particular a {\em Nash Equilibrium} $-$ which is a basic concept which provides the most accepted prediction of a game outcome $-$ of a game with two players is a pair of strategies, one for each player, such that each player has no benefit in independently changing his strategy, i.e., such that these two strategies are mutually optimal responses.


\subsection{Reference \cite{[a12]}}

Reference \cite{[a12]} studies a scenario (see Case I of \cite{[a12]}) in which two hospitals, say hospital I and hospital II, have the same "market", made up of three groups of people: Sportspeople (22\%), Women ( 46\%), Children (32\%). Each hospital has funds to equip reception for only two groups of people. This scenario corresponds to the following game (static with complete information):

:: Players: the set of players is \{hospital I, hospital II\};

:: Strategies: the set of strategies at disposal to each player is \{SW, SB, DB\}, where SW is the strategy to equip reception for Sportspeople and Women, SC for Sportspeople and Children, WC for Women and Children;

:: Payoffs: the payoffs that players get for each combination of strategies are reported in the (bi) matrix in Figure 1; in particular, the authors assume that if hospitals turn to the same group, then hospitals divide that group into equal parts.

\begin{center}
\begin{tabular}{|c||c|c|c|c|c|} \hline
& $II : SW$ & $II : SC$ & $II : WC$
\\ \hline \hline
$I : SW$ & $33\%,33\%$ & $57\%,43\%$ & ${\bf 45\%},{\bf 55\%}$
\\ \hline
$I : SC$ & $43\%,57\%$ & $34\%,34\%$ & $38\%,{\bf 62\%}$
\\ \hline
$I : WC$ & ${\bf 55\%},{\bf 45\%}$ & ${\bf 62\%},38\%$ & $39\%,39\%$
\\ \hline
\end{tabular}
\end{center}

\begin{center}
Figure 1
\end{center}



For example, if I chooses SW and if II chooses SC, then: I and II divide the share of Sportspeople (11\% for each), I gets the share of Women (46\%), II gets the share of Children (32\%); then, at the end, I gets 11\% + 46\% = 57\%, II gets 11\% + 32\% = 43\%.

This game has two Nash Equilibria, i.e., it has two predicted outcomes:

first predicted outcome: I choose SW, II chooses WC;

second predicted outcome: I choose WC, II chooses SW.

\subsection{Local financing as a Game Theory problem}

Let us assume that each hospital, with its portion of the budget, can make one single ward excel. Then let us study local financing as the following game (static with complete information).

Game $G$:

:: Players: the set of players is $\{q_1, q_2,\ldots,q_{|Q|}\}$;

:: Strategies: the set of strategies at disposal to each player is $\{[r_1], [r_2],\ldots, [r_{|R|}]\}$, where $[r_i]$ is the strategy of making ward of type $r_i$ excel (for $ i = 1,\ldots,|R|$);

:: Payoff: the payoffs that players get for each combination of strategies are reported for the sole case $|Q|$ = 2 and $|R| = 2$, in hospital $q_1$ and hospital $q_2$, in the (bi) matrix in Figure 2; in particular, we assume that if more hospitals choose strategy $[r_i]$ (for $ i= 1,\ldots,|R|$), that is, they turn to the same group $P_i$ of the "new market", then these hospitals divide the group $P_i$ according to the population distribution $(a_1, a_2,\ldots, a_{|Q|})$.

\begin{center}
\begin{tabular}{|c||c|c|c|c|c|} \hline
& $q_2 : [r_1]$ & $q_2 : [r_2]$
\\ \hline \hline
$q_1 : [r_1]$ & $|P_1|/a_1,|P_1|/a_2$ & $|P_1|,|P_2|$
\\ \hline
$q_1 : [r_2]$ & $|P_2|,|P_1|$ & $|P_2|/a_1,|P_2|/a_2$
\\ \hline
\end{tabular}
\end{center}

\begin{center}
Figure 2
\end{center}

Consider the following two numerical examples for the case $|Q| = 2$ and $|R| = 2$.

{\em Example 1.}

Assume that: $|P_1|$ = 1000, $|P_2|$ = 400, ($a_1$, $a_2$) = (1/4, 3/4). Then one obtains the game in Figure 3.

\begin{center}
\begin{tabular}{|c||c|c|c|c|c|} \hline
& $q_2 : [r_1]$ & $q_2 : [r_2]$
\\ \hline \hline
$q_1 : [r_1]$ & $250,{\bf 750}$ & ${\bf 1000},400$
\\ \hline
$q_1 : [r_2]$ & ${\bf 400},{\bf 1000}$ & $100,300$
\\ \hline
\end{tabular}
\end{center}

\begin{center}
Figure 3
\end{center}

This game has one Nash Equilibrium, that is one predicted outcome: $q_1$ chooses $[r_1]$, $q_2$ chooses $[r_2]$. \qed


{\em Example 2.}

Assume that: $|P_1| = 1000$, $|P_2| = 4$, ($a_1$,$a_2$) = (1/4, 3/4). Then one obtains the game in Figure 4.				

\begin{center}
\begin{tabular}{|c||c|c|c|c|c|} \hline
& $q_2 : [r_1]$ & $q_2 : [r_2]$
\\ \hline \hline
$q_1 : [r_1]$ & ${\bf 250},{\bf 750}$ & ${\bf 1000},4$
\\ \hline
$q_1 : [r_2]$ & $4,{\bf 1000}$ & $1,3$
\\ \hline
\end{tabular}
\end{center}

\begin{center}
Figure 4
\end{center}

This game has one Nash Equilibrium, that is one predicted outcome: $q_1$ chooses $[r_1]$, $q_2$ chooses $[r_1]$. \qed


Example 1 and Example 2 have two different predicted outcomes: in Example 1, hospitals diversify their choices, in Example 2 hospitals do not diversify their choices. This seems to be linked to the fact that: in Example 1 the "new market" groups are of the {\em same order of magnitude} ($|P_1| = 1000$, $|P_2| = 400$), while in the Example 2 the groups of "new market"are not of the {\em same order of magnitude} ($|P_1| = 1000$, $|P_2| = 4$).

In order to possibly draw general conclusions, we introduce the following assumption, which will allow us to conclude that the tendency of hospitals is to diversify their choices.

{\sc Assumption 1:}

a) the components of population distribution ($a_1$, $a_2$) have the {\em same order of magnitude} [as above in the two examples: ($a_1$, $a_2$) = (1/4, 3/4)];

b) the "new market". is formed by groups having the {\em same order of magnitude} [as above in Example 1: $|P_1|$ = 1000, $|P_2|$ = 400].

{\bf Observation 1}

Assumption 1-(a) seems to be plausible, by definition of (hospitals) district, in the sense that (hospitals) districts are generally defined and saved in a balanced way; Assumption 1-(b) seems to be plausible, by definition of (type of) wards, in the sense that (types of) wards are generally defined and saved in a balanced way. \qed 

Assumption 1 implies [that is can be formalized by] the following inequalities:

$|P_k| > min \{|P_i| / a_j : i = 1, \ldots , |R|; j = 1, \ldots , |Q|\}$     for $k = 1, \ldots , |R|$,	

that for the case $|Q| = 2$ and $|R| = 2$ are:

$|P_2| > min \{|P_i| / a_j :  i = 1, 2; j = 1, 2\}$;  $|P_1| > min \{|P_i| / a_j : i = 1, 2; j = 1, 2\}$,	

that expresses the following condition: each group of the "new market", corresponding to a certain type of ward, estimated on the whole State is more numerous than the smaller subgroup of the "new market", corresponding to another type of ward, estimated on any (hospital) district.

\begin{prop}\label{Proposition 1}
If Assumption 1 holds true, then game $G$:

\begin{itemize}
\item[(i)] has no Nash Equilibrium in which all players choose the same strategy,
\item[(ii)] has a Nash Equilibrium in which not all players choose the same strategy.
\end{itemize}

\end{prop}

{\bf Proof.} Since Assumption 1 holds true, the above inequalities hold true. The proof is only for the case $|Q| = 2$ and $|R| = 2$ [cf. Table 5 below]. For the case $|Q| = 2$ and $|R| > 2$, the proof can be carried out by a similar approach. For the case $|Q| > 2$ and $|R| > 2$ the proof can be carried out by a similar approach but with a greater level of detail, for that, let us omit to report here for brevity and convenience.

\begin{center}
\begin{tabular}{|c||c|c|c|c|c|} \hline
& $q_2 : [r_1]$ & $q_2 : [r_2]$
\\ \hline \hline
$q_1 : [r_1]$ & $|P_1|/a_1,|P_1|/a_2$ & $|P_1|,|P_2|$
\\ \hline
$q_1 : [r_2]$ & $|P_2|,|P_1|$ & $|P_2|/a_1,|P_2|/a_2$
\\ \hline
\end{tabular}
\end{center}

\begin{center}
Figure 5
\end{center}

{\em Proof of (i)}. By contradiction let us assume that $G$ has a Nash Equilibrium in which all players choose the same strategy, say without loss of generality that ($[r_1]$, $[r_1]$) is a Nash Equilibrium. Then one has that $|P_1| / a_2 > |P_2|$ and that $|P_1| / a_1 > |P_2|$. But this contradicts at least one of the above inequalities.

{\em Proof of (ii)}. Let us say that a pair $(x, y)$ of strategies, where $x$ is a strategy of $q_1$ and $y$ is a strategy of $q_2$, is {\em diversified} if $x \neq y$ and is {\em non-diversified} if $x = y$.

Then let us consider the following exhaustive cases.

Case 1: a pair of non-diversified strategies contains an optimal response.

Without loss of generality, let $([r_1], [r_1])$ be a pair of non-diversified strategies containing an optimal response, and let strategy $[r_1]$ be an optimal response of $q_2$ to strategy $[r_1]$ of $q_1$. Then $|P_1| / a_2> |P_2|$. This implies that $|P_1| > |P_2| / a_2$ [in fact, $|P_1| / a_2 > |P_2|$ implies $|P_1| > |P_2|$, which in turn implies that $|P_1| > |P_2| / a_2$] and that $|P_1| / a_1<|P_2|$, since otherwise at least one of the above inequalities would be contradicted. These two inequalities imply that $([r_1], [r_2])$ is a Nash Equilibrium of $G$. Then, in Case 1, (ii) is proved.

Case 2: no pair of non-diversified strategies contains any optimal response.

Let us write $|P_m| = max \{|P_1|, |P_2|\}$, without loss of generality, let $|P_m| = |P_2|$. Then, for each index $i \neq m$, an optimal response to a player's strategy $[r]$ is $[r_m]$ of the other player. On the other hand, by the assumption of Case 2, one has that $([r_m], [r_m])$ does not contain any optimal response. Let $[r_j]$ be an optimal response of $q_1$ to the strategy $[r_m]$ of $q_2$ (without loss of generality): then, by the above, one has that $j \neq m$ and that $([r_j], [r_m])$ is a Nash Equilibrium of $G$. Then, in Case 2, (ii) is proved. \qed \\ 

{\bf Conclusion for local financing}

In the local financing system: if Assumption 1 holds true, according to Observation 1, then hospitals tend to diversify their excellences according to Proposition 1.

\section{Central financing}

The study of central financing is inspired to reference \cite{[a9]}. For that, first let us report a part of reference \cite{[a9]} in which the author studies a Mathematical Programming problem, then let us study central financing as a similar Mathematical Programming problem.

For any reference about Mathematical Programming, in particular about Integer Linear Programming, let us refer to \cite{[Hiller]}.

\subsection{Reference \cite{[a9]}}\label{Section 3.1}

Reference \cite{[a9]} studies the following well-known problem.

{\em Plant Location}: A certain number of plants must be activated, to be chosen over a set of potential plants, which will have to supply a group of customers. Activating the potential plant $j$ involves a cost $d_j$. It is assumed that each facility has unlimited supply capacity. Each customer will be connected to a single plant; in particular, connecting customer $i$ to plant $j$ entails a cost $c_{ij}$. The problem is to choose which plants should be activated in order to minimize the total costs for plants activation and for customers connection.

This problem can be formulated in terms of Integer Linear Programming. For this kind of problem $-$ which belongs to the class of NP-hard problems, see e.g. \cite{[Papadimitriou]}, for an introduction to this kind of problems and for related topics $-$ it becomes difficult to compute an optimal solution, from the point of view of computation, when the problem involves a large number of variables subject to integrality constraints. Then it is usual to determine, instead of an optimal solution of the problem, a heuristic solution of the problem: that is an admissible solution which is computed by a heuristic method, i.e., by a quick and common sense method that provides a presumably good solution (which is not necessarily an optimal solution). In particular, the heuristic solution chosen in \cite{[a9]} for Plant Localization is that produced by a heuristic method called the Greedy Algorithm.

\subsection{Central financing as a Mathematical Programming problem}

Let us study central financing as the following problem.

{\em SGM}: The State has a budget $B$ to make excel in some hospital some wards with the aim to (possibly) hospitalize patients of $P$ (of the "new market"). Making excel in hospital $q$ ward $r$ entails a cost $C_{qr}$. It is assumed that in each hospital each ward has sufficient reception capacity. Every patient of $P$ will be hospitalized either in a made excel ward inside the State or outside the State; in particular, hospitalizing patient $p$ in hospital $q$ in a made excel ward $r$ entails a cost/discomfort $c_{pqr}$ (for the patient), while hospitalizing patient $p$ outside the State entails a cost/discomfort $c_{p,out}$ (for the patient). The problem [of the State] is to choose, in each hospital, which wards should be made excel in order to minimize the overall costs sustained by the State and by patients of $P$.

SGM can be formulated as a problem of Integer Linear Programming as follows: \\

:: Parameters

$B$ = budget available to the State to make excel some hospital wards.

For $p \in P$, for $q \in Q$, for $r \in R$:

$c_{pqr}$ = cost/discomfort for patient $p$ if $p$ is hospitalized in hospital $q$ in the (made excel) ward $r$;

$c_{p,out}$ = cost/discomfort for patient $p$ if $p$ is hospitalized outside the State;

$C_{qr}$ = cost to make excel in hospital $q$ the ward $r$. \\

:: Decision variables

$y_{qr}$ for $q \in Q$, for $r \in R$

$x_{pqr}$ for $p \in P$, for $q \in Q$, for $r \in R$

$x_{p,out}$ for $p \in P$

$y_{qr} = 1$ if in hospital $q$ ward $r$ is made excel;

$y_{qr}  = 0$ otherwise;

$x_{pqr} = 1$ if patient $p$ is hospitalized in hospital $q$ in the (made excel) ward $r$;

$x_{pqr} = 0$ otherwise;

$x_{p,out} = 1$  if patient $p$ is hospitalized outside the State;

$x_{p,out} = 0$ otherwise. \\

:: Model

min	 			$\sum_{q=1}^{|Q|} \sum_{r=1}^{|R|} C_{qr}y_{qr} + \sum_{p=1}^{|P|} \sum_{q=1}^{|Q|} \sum_{r=1}^{|R|} c_{pqr}x_{pqr} + \sum_{p=1}^{|P|} c_{p,out}$  

subject to:	

$x_{p,out}	+ \sum_{q=1}^{|Q|} \sum_{r=1}^{|R|} x_{pqr} = 1$ 	for $p \in P$

(constrains to ensure that: every patient $p$ is hospitalized either in a hospital $q$ in a ward made excel $r$ either outside of the State)
			
$x_{pqr}$ $\leq$ $y_{qr}$ for $p \in P$, for $q \in Q$, for $r \in R$

(constraints to ensure that: in hospital $q$ ward $r$ is made excel if at least one patient $p$ is hospitalized therein)	
				
$\sum_{q=1}^{|Q|} \sum_{r=1}^{|R|} C_{qr}y_{qr} \leq B$

(budget contraint)
				
$x_{pqr} \in \{0, 1\}$		for $p \in P$, for $q \in Q$, for $r \in R$

$y_{qr} \in \{0, 1\}$		for $q \in Q$, for $r \in R$
	
$x_{p,out} \in \{0, 1\}$	for $p \in P$

(integrality constraints \{0, 1\})	\\

Let us observe that, by the above, SGM may admit the following optimal solution:

$x_{pqr}$ = 0 	     for $p \in P$, for $q \in Q$, for $r \in R$

$y_{qr} = 0$		for $q \in Q$, for $r \in R$

$x_{p,out} = 1$	     for $p \in P$

that is an optimal solution such that in each hospital no ward is made excel and all patients in $P$ are hospitalized outside the State.

This could happen for two occurrences:

(i) budget $B$ is insufficient to make in any hospital any ward excel;

(ii) once considered the costs to make in any hospital any wards excel and the costs/discomforts (for patients in $P$) to hospitalize either inside or outside the State, it is not advisable to make in any hospital any ward excel, that is, all $P$ patients are hospitalized outside the State.

In order to avoid the above occurrences let us introduce the following assumption.

{\sc Assumption 2:} to avoid occurrence (i), we assume that $B \geq min\{C_{qr} : q \in Q, r \in R\}$; to avoid occurrence (ii), we assume that the sum of terms ($c_{p,out} - c_{pqr}$) over indices $p, q, r$ is greater than the sum of terms ($C_{qr}$) over indices $q, r$: this assumption is consistent with the fact that one  presumably has $c_{p,out} \geq c_{pqr}$ (for a patient of $P$ it means more cost/discomforts to be hospitalized outside the State rather than within the State) and with the fact that patients of $P$ are numerous.

Let us report below a heuristic method, namely the Greedy Algorithm, to compute a heuristic solution of SGM.

Let $Y = \{y_{qr} : q \in Q, r \in R\}$. A subset $T \subseteq Y$ is {\em admissible} if the sum of values $(C_{qr})$ over indices $q$ and $r$ such that $y_{qr} \in T$ is less than or equal to $B$ [i.e. if the budget constraint is satisfied].
By definition of the SGM problem, each admissible solution of the SGM problem is associated with an admissible subset $T \subseteq Y$. In fact: on the one hand, an admissible solution clearly identifies an admissible $T \subseteq Y$ subset; on the other hand, once fixed an admissible subset $T \subseteq Y$, it remains automatically defined a "best possible solution" associated with $T$, which is the solution in which patients in $P$ will be hospitalized to a destination [either among those in $T$, or outside the State] which entails a lowest cost/discomfort; in particular, for each admissible subset $T \subseteq Y$, we denote as $Z (T)$ the value of the objective function for a "best possible solution" associated with $T$.

Thus SGM can be re-defined as follows:

{\em SGM}: Determine an admissible subset $T \subseteq Y$ such that $Z(T)$ is minimum.

Let us report from \cite{[a9]}, with appropriate modifications, the following Greedy Algorithm to compute an admissible subset $T \subseteq Y$. The Greedy Algorithm is a heuristic method, that is, a quick and common sense method provides a presumably good solution (which is not necessarily an optimal solution), i.e., an admissible solution $T$ such that $Z(T)$ is presumably not far from the optimal solution value of SGM. \\

{\bf Greedy Algorithm}

Inizializzation: $i$ = 1, $T_0 = \emptyset$, $Z(T0) = - \infty$.

{\em begin of the $i$-th iteration}

$y_i = argmin \{Z(T_{i-1}) \setminus \{y\} : y \in Y \ T_{i-1}\}$.

If $Z(T_{i-1} \setminus \{y_i\}) \geq Z(T_{i-1})$, then STOP: $T_{i-1}$ is the greedy solution.

Otherwise set $T_i = T_{i-1} \setminus \{yi\}$.

If $T_i= Y$, then STOP: $T_i$ is the greedy solution.

Otherwise set $i = i+1$ and go to the $i$-th iteration.

{\em end of the $i$-th iteration} \\

Let us report from \cite{[a9]}, with appropriate modifications, the following observations on the Greedy Algorithm: once an element $y \in Y$ is inserted in the (current) subset $T \subseteq Y$, then it is no longer removed; the element to be inserted in the (current) subset $T \subseteq Y$ is chosen, at each step, by looking just at the immediate advantage consisting in a decrease in the value of the objective function.

Let us introduce the following assumption:

{\sc Assumption 3:} we accept the greedy solution of SGM produced by the Greedy algorithm as the solution to SGM even if it is not guaranteed to be an optimal solution, according to the above and to the fact that SGM involves a large number of variables subject to integrality constraints.

Then in order to possibly draw general conclusions [on the structure of the greedy solution] let us also introduce the following assumptions.

{\sc Assumption 4:} cost/discomfort $c_{pqr}$ depends only on the district in which $p$ is located and on the distance from this district to district $q$ [i.e., it does not depend on $r$]; in particular, this cost is the same for every person in a given district.

{\sc Assumption 5:} $C_{qr} = K$, for each $q \in Q$, for each $r \in R$; that is, the cost to make excel in hospital $q$ ward $r$ is a standard cost $K$ that does not depend on $q$ and $r$.		

{\bf Observation 2}

Assumption 2 seems to be plausible since it is consistent with plausible facts. Assumption 3 seems to be acceptable according to what is reported in Subsection \ref{Section 3.1}. Assumption 4 seems to be plausible at least in the general case. Assumption 5 is expressed in an improbable manner, however it seems to be plausible that all these costs have the same order of magnitude: then, in this approximate context where the greedy solution is accepted as the solution to the SGM problem, Assumption 5 seems to be plausible. \qed 

{\bf Observation 3}

It is possible to define two total orders, one for set $R$, one for set $Q$.

:: The first total order refers to set $R = \{r_1,r_2, \ldots ,r_{|R|}\}$ of (types of) wards according to the baseline scenario introduced above [cf. (1.7)].

That is: we assume that $r_1,r_2, …, r_{|R|}$ are ordered so that $|P_1| \geq |P_2| \geq \ldots \geq |P_{|R|}|$.

:: The second total order refers to set $Q = \{q_1,q_2, \ldots ,q_{|Q|}\}$ of hospitals according to Assumptions 4-5.

That is, once fixed a (type of) ward $r$, we assume that $q_1,q_2, \ldots ,q_{|Q|}$ are ordered so that:

$q_1$ = hospital such that, wishing to make excel in exactly one hospital the ward $r$, then the most convenient hospital is hospital $q_1$;

$q_2$ = hospital such that, wishing to make excel in exactly one hospital the ward $r$, assuming that in hospital $q_1$ ward $r$ is already made excel, then the most convenient hospital is hospital $q_2$;

more generally, for $k \geq 2$, we assume that:

$q_k$ = hospital such that, wishing to make excel in exactly one hospital the ward $r$, assuming that in hospitals $q_1, \ldots, q_{k-1}$ ward $r$ is already made excel, then the most convenient hospital is hospital $q_k$.

This order is the same for all (types of) wards by Assumptions 4-5.  \qed 

{\bf Observation 4}

The total orders $r_1,r_2, \ldots ,r_{|R|}$ and $q_1,q_2, \ldots, q_{|Q|}$ defined in Observation 3 can be easily computed once the parameter values are defined. \qed 

{\bf Observation 5}

The total order $q_1,q_2,\ldots,q_{|Q|}$ of hospitals (of Observation 3) does not necessarily coincide with the total order of hospitals that would be obtained on the basis of the population in the respective districts. For example, in Abruzzo (Italy) the total order of hospitals on the basis of the population of the respective districts is presumably: Pescara, Chieti, L'Aquila, etc., while the total order $q_1,q_2,\ldots,q_{|Q|}$ of hospitals (of Observation 3) is presumably: Pescara, L'Aquila, Chieti, etc. This is since, once a ward $r$ is made excel at Pescara, then (from the point of view of the minimization of costs) according to Assumption 4 it could be agreed that the next ward $r$ is made excel at L'Aquila instead of at Chieti although the population of (the district of) Chieti is greater than the population (of the district) of L'Aquila.  \qed 

\begin{prop}\label{Proposition 2}
If Assumptions 4-5 hold true, then the greedy solution of the SGM problem is such that:

if $y_{qr} = 1$, then $y_{q'r'} = 1$ for every $q'$ that precedes $q$ in the total order $q_1,q_2,\ldots,q_{|Q|}$ defined in Observation 3, and for every $r'$ that precedes $r$ in the total order $r_1,r_2,\ldots,r_{|R|}$ defined in Observation 3.
\end{prop}

{\bf Proof}. The proof derives directly by definition of the Greedy Algorithm and by definition of the total orders in Observation 3.	\qed \\ 

In other words, the greedy solution of the SGM problem is such that, with reference to the total orders defined in Observation 3, if in hospital $q_h$ ward $r_k$ is made excel, then in hospital $q_i$, for $i = 1, 2, ..., h$, (at least) wards $r_1,r_2, \ldots, r_k$ are made excel.

For example, if $|Q| = 5$ and if $|R| = 7$, then the greedy solution of the SGM problem is of the form:

in $q_1$: wards $r_1,r_2,r_3,r_4,r_5$ are made excel

in $q_2$: wards $r_1,r_2,r_3,r_4$ are made excel

in $q_3$: wards $r_1,r_2$ are made excel

in $q_4$: no ward is made excel

in $q_5$: no ward is made excel \\

{\bf Conclusion for central financing}

In the central financing system: if Assumptions 2-5 hold true, according to Observation 2, then the State tends to create poles of excellence according to Proposition 2. \\


{\bf Acknowledgements.} Please I would like to thank Prof. Fausto Di Vincenzo for having explained to me his paper \cite{[a7]}. Then please I would like to witness that just try to pray a lot and am not able to do anything without that - ad laudem Domini.

\begin{footnotesize}

\end{footnotesize}

\end{document}